\documentclass{iau}
\usepackage{graphicx}
\usepackage{natbib}
\usepackage{aas_macros}

\newcommand{\pkstwo}{{PKS\,0023$-$26}}
\newcommand{\pksone}{{PKS\,1549$-$79}}

\newcommand{\kms}{$\,$km$\,$s$^{-1}$}
\newcommand{\WHz}{$\,$W$\,$Hz$^{-1}$}

\newcommand{\msun}{{$M_\odot$}}
\newcommand{\msunyr}{{$M_\odot$ yr$^{-1}$}}

\newcommand{\co}{{CO(1-0)}}

\def\HI{H{\,\small I}}

\def\emph#1{{\sl #1}}
\newcommand{\ltsima} {$\; \buildrel < \over \sim \;$}
\newcommand{\gtsima} {$\; \buildrel > \over \sim \;$}
\newcommand{\lta} {\lower.5ex\hbox{\ltsima}}
\newcommand{\gta} {\lower.5ex\hbox{\gtsima}}

\title[IAU359.~~Jet-ISM interplay with ALMA] 
{Taking snapshots of the jet-ISM interplay with ALMA
 }

\author[Morganti et al.]   
{Raffaella Morganti$^{1,2}$, Tom Oosterloo$^{1,2}$,  Clive N. Tadhunter$^3$
}

\affiliation{$^1$ASTRON, the Netherlands Institute for Radio Astronomy, Oude Hoogeveensedijk 4,\\ 7991PD Dwingeloo, The 
Netherlands.  email: {\tt morganti@astron.nl} \\[\affilskip]
$^2$Kapteyn Astronomical Institute, University of Groningen, Postbus 800,\\
9700 AV Groningen, The Netherlands \\
$^3$ Department of Physics and Astronomy, University of Sheffield, Sheffield, S7 3RH, UK
}

\pubyear{2020}
\volume{359}  
\jname{Galaxy evolution and feedback across different environments}
\editors{T. Storchi-Bergmann, R. Overzier, W. Forman \& R. Riffel, eds.}

\begin{document}

\maketitle

\begin{abstract}
We present an update of our on-going project to characterise the impact of radio jets on the ISM. This is done by tracing the distribution, kinematics and excitation of the molecular gas at high spatial resolution using ALMA. The radio active galactic nuclei (AGN) studied are in the interesting phase of having a  recently born radio jet.  In this stage, the plasma jets can have the largest impact on the ISM, as also predicted by state-of-the-art simulations. The two targets we present  have quite different ages, allowing us to get snapshots of the effects of  radio jets as they grow and evolve. Interestingly,  both also host powerful quasar emission, making them ideal for studying the full impact of  AGN.
The largest mass outflow rate of molecular gas is found in a  radio galaxy (\pksone) hosting a newly born radio jet still in the early phase of emerging from an obscuring cocoon of gas and dust. Although the molecular mass outflow rate is high (few hundred \msunyr), the outflow is limited to the inner few hundred pc region. \\
In a second object (\pkstwo), the jet is larger (a few kpc) and is in a more advanced evolutionary phase.  In this object, the distribution of the molecular gas is reminiscent of what is seen, on larger scales, in cool-core clusters hosting radio galaxies. 
Interestingly, gas deviating from quiescent kinematics (possibly indicating an outflow) is not very prominent, limited only to the very inner region, and has a low mass outflow rate. Instead, on kpc scales, the radio lobes appear associated with  depressions  in the distribution of the molecular gas. This suggests that the lobes have  broken out from the  dense nuclear region.  However, the AGN does not appear to be able at present  to  stop the star formation observed in this galaxy. 
These results support the idea that the effects of the radio source start  in the very first phases by producing outflows which, however, tend to be limited to the kpc region. After that, the effects turn into producing large-scale bubbles which could, in the long term, prevent the surrounding  gas from cooling.  
Thus, our results provide a way to characterise the effect of radio jets in different phases of their evolution and in different environments, bridging the studies done for radio galaxies in clusters.

\end{abstract}

\firstsection 
\section{Introduction}
The evolution of massive galaxies appears to be strongly influenced by the energy released during the active phase of their super massive black hole (SMBH). This process, known as feedback, is considered the one regulating (and quenching) their star formation  (e.g.\ \citealt{Harrison17}). Feedback is believed to work in two main modes, both aimed at reducing the amount of cold gas and the related star formation: ``quasar" mode, with gas outflows driven by the active galactic nucleus (AGN), clearing the gas from the host galaxy, and the ``maintenance" mode, where the energy released by the AGN prevents the cooling of the gas on larger scales, from the hot halo  or from the intergalactic medium.  In the commonly assumed picture of AGN  feedback, the role of radio jets is considered to be mostly connected to the latter mode (see e.g.\ \citealt{McNamara12}). This mode complements the effect of outflows/winds considered to dominate in radiatively efficient AGN. 
However, the picture we are getting from the growing number of detailed observations tracing multiple phases of the gas appears to be more complex and these two modes appear intertwined. 
An example of such complexity is the fact that radio jets can also  produce gaseous outflows, thus having an impact on (sub) kpc-scales.  Furthermore,  their relative importance may even change during the life of the AGN.
{\sl Thus, radio AGN with jets represent ideal objects to trace the feedback while they evolve from sub-kpc to many tens of kpc scales. }

The impact of jets in producing outflows has been observed in an increasing number of both high- and low-power (including radio quiet) radio sources (\citealt{Morganti20} and refs therein).
Particularly interesting is that jet-driven outflows are more prominent when the jets are in their initial phase (see e.g.\ \citealt{Holt09,Shih13,Morganti18,Molyneux19} and many others). 
Newly born or young radio jets can be identified thanks to the characteristics of their radio emission (e.g.\ size, and peaked radio  spectrum, for more details see  e.g.\ \citealt{ODea98,Orienti16}). 
The impact of jets in their starting phase is also predicted by numerical simulations. The work of \citet{Wagner12,Bicknell18,Mukherjee16,Mukherjee18} has shown the strong coupling of the newly born jet with the surrounding clumpy inter-stellar medium (ISM). Furthermore, this interaction is predicted to produce a cocoon of shocked gas expanding perpendicular to the jet, thus impacting a much larger volume of the host galaxy than just the region of the jets themselves.

Finally,  the {\sl impact of jets can be dominant even when a radiatively efficient AGN is present}.
The best illustration of this is the case of IC~5063, a Seyfert 2 galaxy, with strong emission lines. This  galaxy hosts low radio power jets  ($P_{\rm 1.4~GHz}\sim 10^{23.4}$ \WHz). Despite the low radio power, they provide one of the clearest examples of jet-induced outflows, where the radio plasma is disturbing the kinematics of {\sl all the phases of the gas} \citep[see][for an overview]{Tadhunter14,Morganti15}. The region co-spatial with the radio emission is where the most kinematically disturbed ionised, molecular and \HI\ gas is located (with velocities deviating up to 600 \kms\ from regular rotation). 
This  jet-ISM interaction is also affecting the  physical conditions of the gas  \citep{Oosterloo17}. 
The properties observed have been  well reproduced by   hydrodynamic simulations and the details of the comparison between the data and the simulation is presented in \citet{Mukherjee18}.  

All this strongly indicates the relevance of radio jets,  particularly  in the initial phases of their evolution. However, after 1-2 kpc (typically after a few Myr) the jet breaks out from the dense central core and the way it interacts with the surrounding medium changes. 
{\sl Thus, should we expect that the impact will change with  jet evolution?} 
This question is particularly important, also because there appears to be a general consensus that most  observed outflows (regardless their origin) are largely limited to the central kpc region while only a small fraction of the outflowing gas is actually leaving the galaxy. Thus, gas outflows may not be enough to supply the required feedback  from AGN. 
Here we focus on two powerful radio sources  ($P\sim 10^{26}-10^{27}$ \WHz) with jets in different phases of their {\sl initial} evolution. Interestingly, the sources  also host radiativly efficient AGN. Thus, in these objects there is no shortage of energy released by their active SMBH.  We use molecular gas as tracer of the impact of the AGN because it is typically found to carry most of the outflowing mass. This is part of a larger project to understand the impact of radio jet as function of their properties (i.e.\ power, age, environment etc.; see also \citealt{Maccagni18,Oosterloo17} for other objects studied).
 
\begin{figure}
\begin{center}
 \includegraphics[width=10cm]{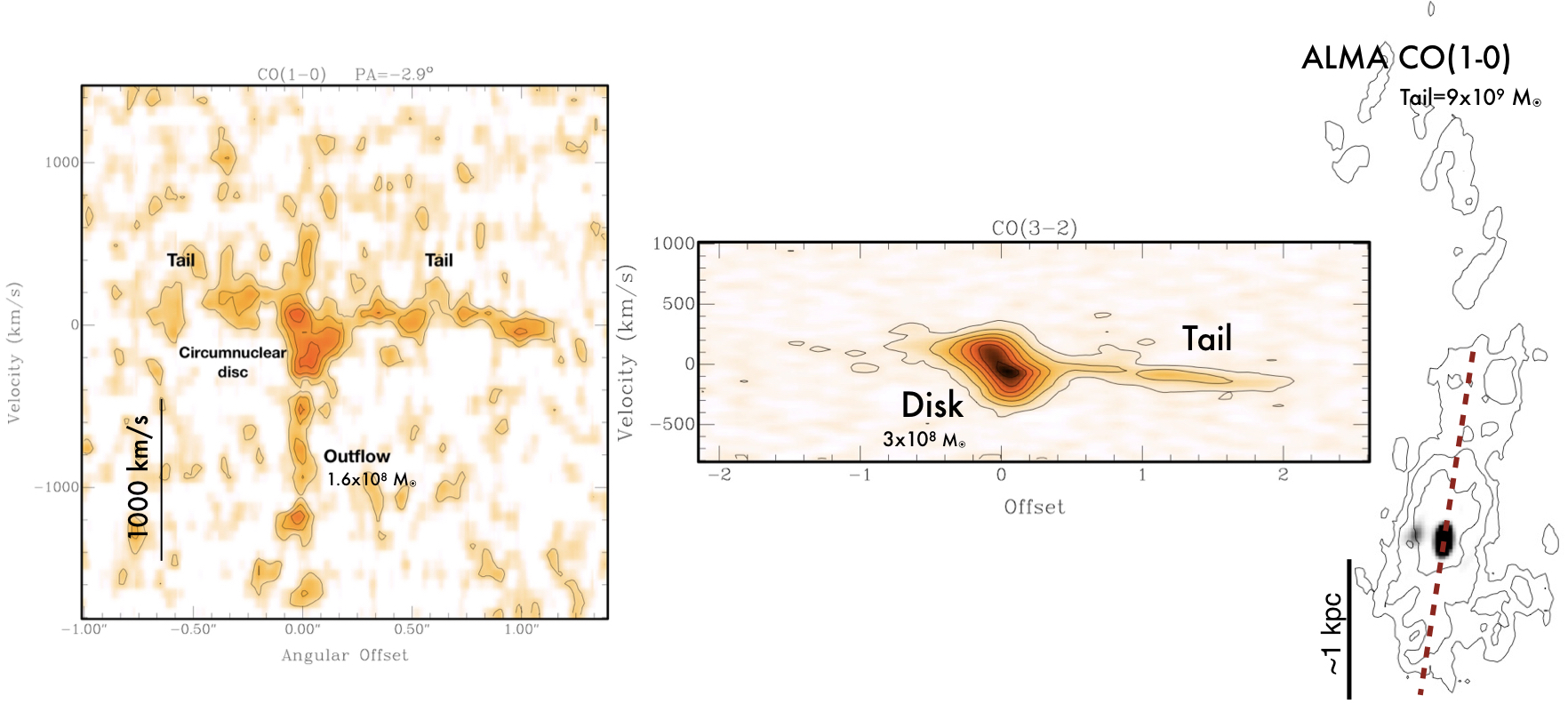} 
 \caption{The distribution and kinematics of the molecular gas in  the central few hundred pc of \pksone\ as obtained with ALMA, see \cite{Oosterloo19} for details. ALMA CO(1-0) and CO(3-2) detected in emission with spatial resolution ranging from 0.05 arcsec ($\sim 100$ pc) to 0.2 arcsec. The position-velocity plots are made along the dashed line, \cite{Morganti20}.}
   \label{fig:fig1}
\end{center}
\end{figure}

\section{Accretion and feedback in an obscured, young radio quasar}

The first object shown here is \pksone\  ($z = 0.150$), having a young radio jet ($\sim 300$ pc in size) hosted by an obscured, far-IR bright quasar and, therefore, in a particularly crucial early stage in the evolution \citep{Holt06}. As expected, a fast outflow of warm ionised gas as well as a Ultra Fast Outflow in X-rays  \citep{Tombesi14} are present, but the kinetic power of the warm outflow $\sim 4\times10^{-4}\, L_{\rm edd}$.

The \co\ and CO(3-2) ALMA high resolution observations (\citealt{Oosterloo19}) show the presence of  three gas structures, which can be seen in Fig.\ \ref{fig:fig1}.  Kiloparsec-scale tails are observed, resulting from an on-going merger which  provide gas that is accreting  onto the centre of PKS~1549--79. At the same time, a circum-nuclear disc has formed in the inner few hundred parsec, and a very broad ($> 2300$ \kms) component associated with fast outflowing molecular gas is detected at the position of the AGN. As expected, the outflow is massive ($\sim$600 \msunyr) but, despite the fact that \pksone\ should represent an ideal case of feedback in action, it is limited to the inner 200 pc. These results illustrate that the impact on the surrounding medium of the energy released by the AGN is not always as expected from the feedback scenario. The outflow of warm, ionised gas is slightly more extended (see \citealt{Oosterloo19}), but  modest in terms of mass outflow rate ($\sim 2$ \msunyr; \citealt{Holt06}, Santoro et al. in prep). 

Both the jet and the radiation could drive the outflow. Circumstantial evidence suggests that the jet may play the prominent role. We observed a strongly bent component of the jet, characterised by a very steep radio spectrum. This suggests that this part of the jet is a remnant structure, possibly resulting from a strong interaction that has temporarily destroyed the jet.  This interaction could have originated  the massive outflow. This interaction is also impacting the conditions of the gas, as seen from the high ratio CO(3-2)/CO(1-0) found in the central regions. The depletion time is relatively short ($10^5 -10^6$ yr), suggesting that the outflow will last only for a relatively short time.

 \section{\pkstwo: a few kpc-scale young radio AGN}

\begin{figure}
\begin{center}
 \includegraphics[width=10cm]{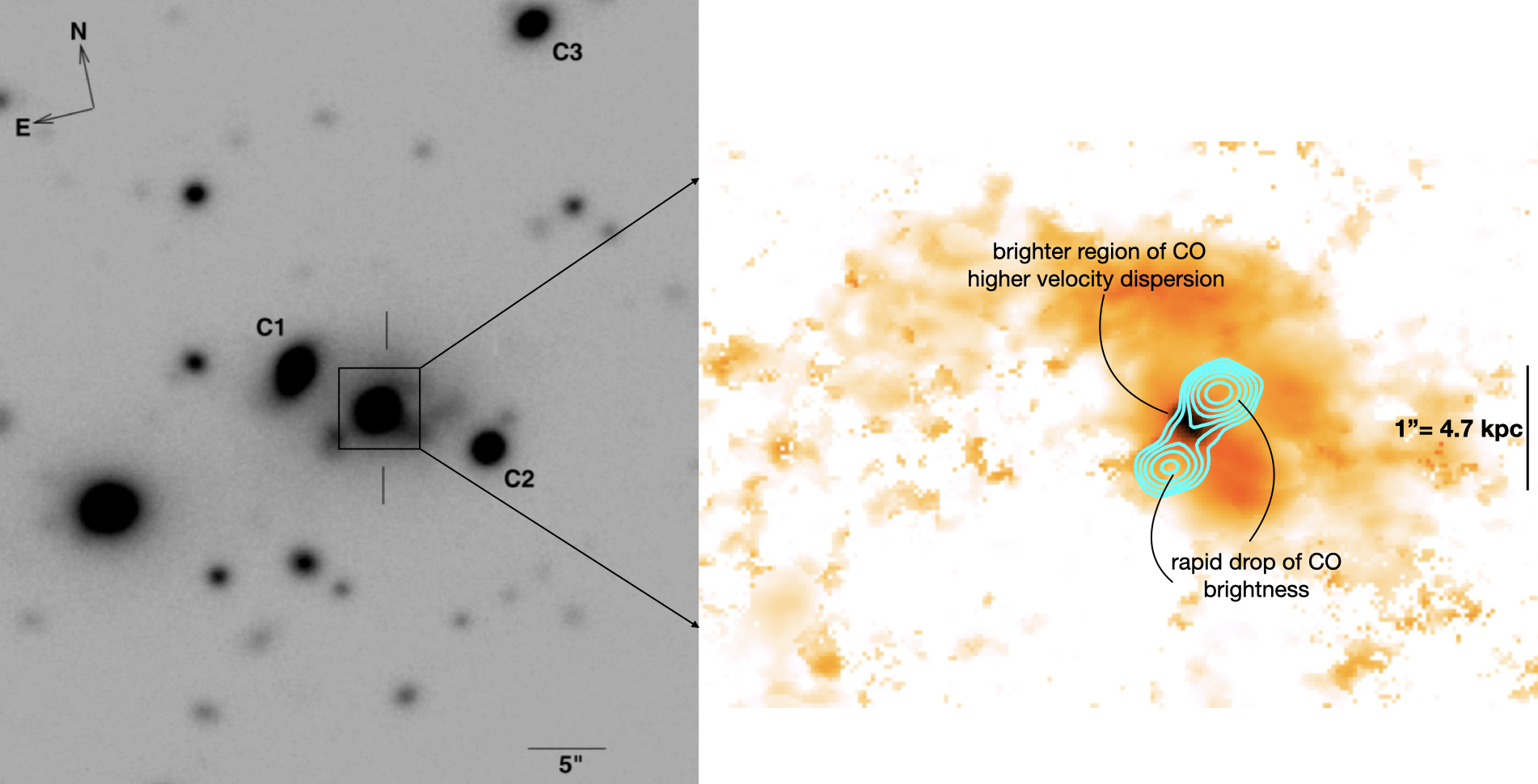} 
 \caption{{\bf Left} Optical  image  from  Gemini GMOS-S. Marked as C1, C2, C3 galaxies confirmed to have redshifts similar to \pkstwo. More objects and tails are seen even closer ($\sim 10$ kpc) to the target galaxy, \cite{Ramos13}; {\bf Right} Total intensity of the molecular gas (orange scale) with superimposed the contours (cyan) of the continuum emission. The center(core) of the radio emission is coincident with the peak of the molecular gas.}
   \label{fig:fig2}
\end{center}
\end{figure}

A second object, \pkstwo\ ($z = 0.32188$), was selected because, although  still a  young radio source,  it is in a more evolved  phase having reached already a few kpc in size. Perhaps unusual for powerful radio galaxies, it is located in a dense environment (see Fig.\   \ref{fig:fig2}, left; \citealt{Ramos13}). Also interesting is the presence in the host galaxy of an extended region with a very young stellar population ($\sim 30$ Myr, \citealt{Holt06,Tadhunter11}). The corresponding star formation rate ($\sim 30$ \msunyr) is consistent  with what is expected for main-sequence star forming galaxies of similar stellar mass as the host galaxy.
The deep, high resolution (0.2 arcsec) ALMA  CO(2-1)  observations reveal that \pkstwo\ is embedded in $5\times10^{10}$ \msun\ of molecular gas, distributed over about 20 kpc (see Fig.\  \ref{fig:fig2}, right).
Interestingly, the distribution is reminiscent of those seen in cool-core clusters (e.g.\  \citealt{Russell19}), because it appears offset from the centre of the galaxy (and radio source). Part of the gas distributed in filaments with relatively smooth velocity gradients reaching out some of the companion galaxies. However, the large amount of molecular gas detected, and the high velocity dispersions observed, suggest that, at least part of the gas is  coming from the cooling of the hot X-ray halo (tentatively detected with XMM; \citealt{Mingo14}).  

The central region is  brightest in CO (Fig.\ \ref{fig:fig2}, right), either because the molecular has gas piled up there, or because  it has higher excitation due to the stronger impact of the AGN (as seen in other objects; \citealt{Oosterloo17,Oosterloo19}). Indeed, in this object the gas with velocities deviating from the quiescent kinematics is also located in the central sub-kpc region. However, the velocities are low (not more than $\sim 300$ \kms). If associated with an outflow, the mass outflow rate is much more modest than in \pksone. Interestingly, this appears to follow the trend found by \cite{Holt08} for the ionised gas: the amplitude of the outflows decreases as the radio jets expand. 
Outside this central region, the brightness of the molecular gas drops rapidly in the regions of the radio lobes. 
A possible explanation is that the jets  have already broken out from the dense, central region and they are now starting to create thermal bubbles which, at a certain point time, will provide a way to prevent the cooling of the hot ISM.  
Thus, the AGN (optical and radio) does not have {\sl at present} any substantial impact on the gas on tens kpc galaxy scales where substantial star formation is  ongoing from the large reservoir of molecular gas.
Possibly, the young radio source is still too young and is in an early phase of interaction with the rich gaseous medium and  {\sl only starting} to affect it. 

\section{Connecting the two objects: evolution of the impact of the jets?}

Based on the results on these two objects (and another handful studied in detail in literature), we suggest that in the first phases (i.e.\ in the sub-kpc region and for ages $< 10^6$yr) the radio jets are expanding in the inner dense, clumpy ISM where the coupling between the jet and the ISM is very strong (e.g.\ \citealt{Mukherjee18}). In this phase, they can drive fast and massive outflows. These can inflate a  cocoon of gas expanding perpendicular to the jet, shocking and kinematically disturbing the gas.  
Although the mass outflow rate of the molecular gas can be large in very young jets (as found in \pksone), the size of the region affected can be  limited to a few hundred pc. Furthermore, the speed of the outflows appears to decrease as the jet expands as seen in \pkstwo\ (and found for the ionised gas; \citealt{Holt09}).

However, the impact appears to change as the jet evolves. When the radio jets expand further, i.e.\  outside the 1-2 kpc region, they break out from the dense central gas and the type of interaction changes, becoming more similar to the one observed e.g.\ in X-rays with the formation of cavities and jet-driven expanding bubbles in the host galaxy and in the IGM.  

In the case of \pkstwo, the gas at kpc scales is still forming stars, unaffected by the AGN. This will likely continue until the available molecular gas is depleted by this process ($\sim 10^9$ yr) while in the meantime the effect of the growing jets will likely increase.  When they reach larger scales (tens of kpc on time scales of a few $\times 10^7 - 10^8$ yr), they can become more efficient in preventing more of the hot gas in the halo from cooling and, therefore, quenching  future star formation. It interesting that studies focusing on the star formation rate and AGN luminosity are reaching similar conclusions (see \citealt{Harrison19}).
In order to confirm the trends found so far and to test this scenario, we need to expand the number of radio galaxies studied, while covering a large parameter space in terms of age and radio power, as well as exploring the properties of the hot gas.

\end{document}